\documentclass[reprint,
superscriptaddress,
amsmath,amssymb,aps,showkeys,showpacs,
twoside,final,secnumarabic,
nofootinbib]{revtex4-2}

\usepackage[paperwidth=205mm,paperheight=290mm,top=17mm,bottom=25mm,
inner=17mm,outer=17mm,
twoside]{geometry}

\usepackage{cmap} 
\usepackage[T1,T2A]{fontenc}
\usepackage[utf8]{inputenc}
\usepackage[russian,english]{babel}
\usepackage{color}
\usepackage{graphicx}
\usepackage{dcolumn}
\usepackage{bm} 
\usepackage[unicode=true,colorlinks=true,linkcolor=magenta, urlcolor=blue, citecolor = blue,breaklinks]{hyperref}
\usepackage{multirow}
\usepackage{url}
\usepackage{breakurl}
\DeclareGraphicsExtensions{.eps}

\newcount\issue
\newcount\Vol
\newcount\numb
\usepackage{fancyhdr} 
\def\Vol{\textbf{78}}
\def\numb{x}
\begin{document}

\title{Neutrino Physics\\[20pt]
Expected flavor composition of supernova neutrinos} 

\def\addressa{address 1}
\def\addressb{address 2}

\author{\firstname{Antonio}~\surname{Capanema}}
\affiliation{Gran Sasso Science Institute (GSSI),
Viale Francesco Crispi 7, L'Aquila, Italy}
\affiliation{INFN - Laboratori Nazionali del Gran Sasso (LNGS),
Via Giovanni Acitelli 22, L'Aquila, Italy}
\author{\firstname{Yago}~\surname{Porto}}
\email[E-mail: ]{yago.porto@tum.de }
\affiliation{Technische Universit\"at M\"unchen,
James-Franck-Str.~1, 85748 Garching, Germany}
 \author{\firstname{Maria Manuela}~\surname{Saez}}
\affiliation{RIKEN iTHEMS, 2-1 Hirosawa, Wako, Saitama
351-0198, Japan}
\affiliation{Department of Physics, University of California, Berkeley, California 94720, USA}

\received{xx.xx.2025}
\revised{xx.xx.2025}
\accepted{xx.xx.2025}

\begin{abstract}
We revisit the flavor composition of neutrinos from core-collapse supernovae (SN),
focusing on robust predictions that are insensitive to the poorly known dynamics of
collective flavor conversion in the inner core. Assuming that the many different
trajectories and microscopic histories of neutrinos lead to decoherence of the
ensemble at the boundary between the region of collective effects and the
Mikheyev-Smirnov-Wolfenstein (MSW) dominated layers, we show that standard
matter effects alone strongly constrain the electron-flavor fraction at Earth.
For normal mass ordering (NO) we obtain $f_{\nu_e}^{\rm NO}\lesssim 0.5$ at all
times and energies, while for inverted ordering (IO), we predict
$f_{\nu_e}^{\rm IO}\simeq 1/3$, i.e.\ near flavor equipartition. Shock-wave
propagation through the high (H) MSW resonance drives the system toward
equipartition also in NO. In this way our framework links simple
assumptions about decoherence and standard matter effects to robust expectations
for the flavor evolution inside core-collapse supernovae. This contribution
summarizes the main results of Ref.~\cite{Capanema:2025}.
\end{abstract}

\keywords{supernova neutrinos; flavor composition; flavor conversion; MSW effect   \\[5pt]}

\maketitle
\thispagestyle{fancy}


\section{Introduction}\label{intro}

Core-collapse supernovae release about $10^{53}\,\text{erg}$ of binding energy,
$\sim 99\%$ of which is carried away by neutrinos over $\mathcal{O}(10)\,\text{s}$~\cite{Janka:2017}.
Only a handful of events from SN~1987A were detected, yet they were
broadly consistent with, and provided important evidence for, the basic
picture of neutrino-driven explosions. The next Galactic SN will be
observed with large detectors and will provide a high-statistics neutrino signal.
However, interpreting this signal is challenging because flavor evolution in the
dense inner regions is dominated by $\nu$-$\nu$ interactions, leading to
nonlinear collective effects whose outcome is still under debate~\cite{Duan:2010, Tamborra:2021}.

\begin{figure}[t]
  \centering
  \includegraphics[width=\columnwidth]{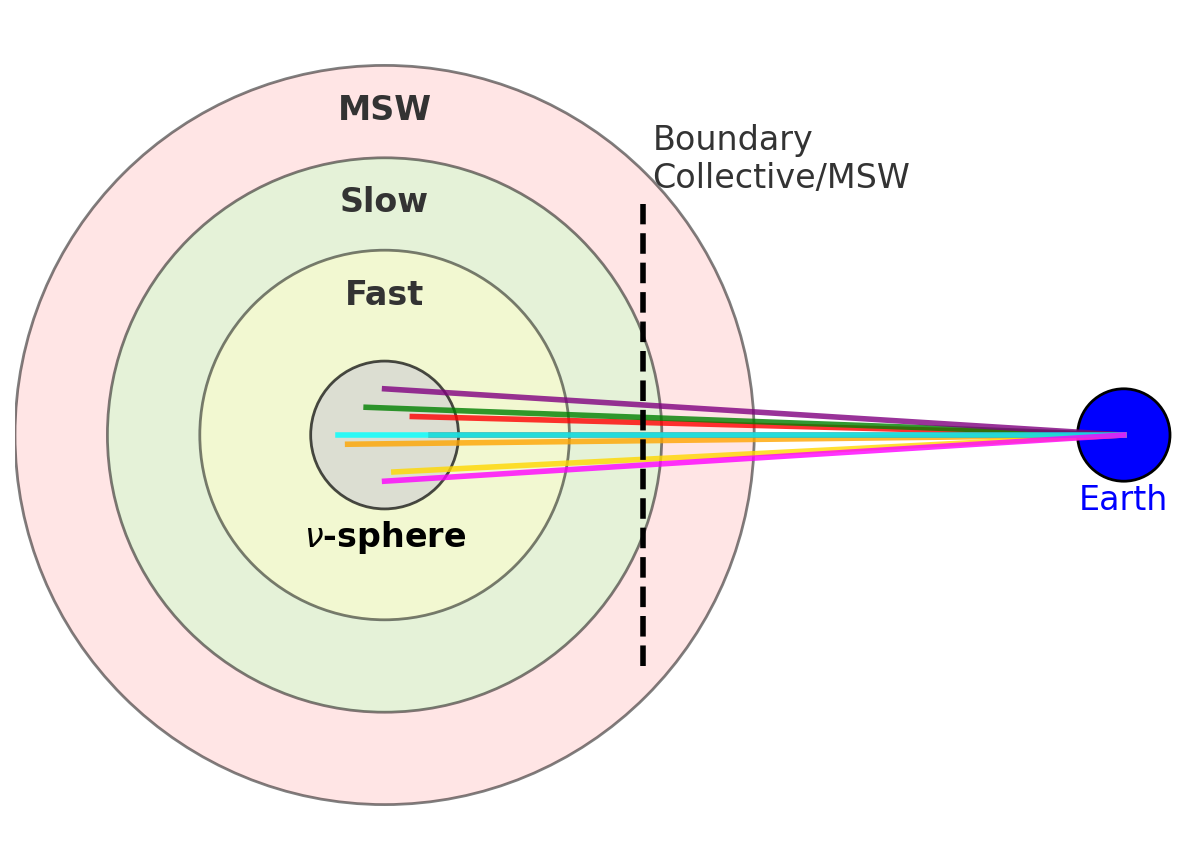}
  \caption{Schematic view of flavor evolution zones in a core-collapse
  supernova. Neutrinos are produced near the $\nu$-sphere, traverse regions
  where fast and slow collective effects can occur, and eventually enter the
  MSW-dominated outer layers. Different trajectories experience different
  histories in the inner regions, which motivates treating the ensemble as
  decohered when it reaches the boundary with the MSW region.}
  \label{fig:layers}
\end{figure}

In this contribution, we summarize the main results of Ref.~\cite{Capanema:2025}, where we asked to what extent the final
flavor composition at Earth can be constrained using only standard matter effects
in the outer layers and a minimal assumption about the state of the neutrino
ensemble after collective conversions. Our strategy is sketched in
Fig.~\ref{fig:layers}: neutrinos undergo complicated evolution in the inner
``fast'' and ``slow'' conversion regions, but we assume that by the time they
enter the MSW layer, phase coherence among different trajectories has been lost.
We then follow the adiabatic evolution through the MSW resonances and derive
general bounds on the electron-flavor fraction at Earth.

\section{\label{sec:level1}Flavor evolution in the outer supernova layers}

Beyond a few hundred kilometers from the $\nu$-sphere, $\nu$-$\nu$
interactions become subdominant and flavor evolution is governed by vacuum
oscillations and coherent forward scattering on the ordinary matter background.
In the flavor basis the Hamiltonian for neutrinos is
\begin{equation}
H_\nu = \frac{1}{2E}\,U\,
\mathrm{diag}(0,\Delta m^2_{21},\Delta m^2_{31})\,U^\dagger
+ \mathrm{diag}(V_e,0,0)\,,
\label{eq:Hnu}
\end{equation}
where $U$ is the PMNS matrix, $E$ the neutrino energy, and
$V_e=\sqrt{2}\,G_F n_e$ the charged-current matter potential.
The evolution is described by $i\,\mathrm{d}\rho_\nu/\mathrm{d}t=[H_\nu,\rho_\nu]$
for the flavor density matrix $\rho_\nu$.

As the matter density decreases, neutrinos cross two MSW resonance layers.
The ``H'' resonance, associated with $\Delta m_{31}^2$ and mixing angle
$\theta_{13}$, occurs at higher density, while the ``L'' resonance involves
$\Delta m_{21}^2$ and $\theta_{12}$. Adiabaticity at a resonance is controlled
by the usual parameter
\begin{equation}
\gamma = \frac{\Delta m^2\,\sin^22\theta}{2E\cos2\theta}
\left|\frac{\mathrm{d}\ln(V_e)}{\mathrm{d}r}\right|^{-1}_{\!\rm res}\!,
\end{equation}
and flavor conversion is adiabatic for $\gamma\gtrsim1$.
In the following we assume adiabatic propagation through the L resonance and
consider both adiabatic and nonadiabatic propagation at the H resonance, which
can be affected by shock waves.

At densities well above the H resonance the matter potential dominates and
$\nu_e$ effectively coincides with the heaviest matter eigenstate, while the
$\nu_\mu$ and $\nu_\tau$ experience nearly identical potentials and form an
approximately maximally mixed doublet. This near $\mu$-$\tau$ symmetry plays
a central role in our argument.

\section{Decoherence and suppression of off-diagonal terms}
Consider the ensemble of neutrinos crossing the boundary between the region of
collective effects and the MSW region in a given time interval. In the flavor
basis the density matrix can be written as
\begin{equation}
\rho_\nu^{(F)} = \frac{1}{N}\sum_{k=1}^N
\begin{pmatrix}
\alpha_k^2 & 0 & 0 \\
0 & \beta_k^2 & \beta_k\gamma_k e^{-i\varphi_k} \\
0 & \beta_k\gamma_k e^{+i\varphi_k} & \gamma_k^2
\end{pmatrix},
\end{equation}
where $\alpha_k^2$, $\beta_k^2$, and $\gamma_k^2$ are the probabilities for the
$k$-th neutrino to be detected as $\nu_e$, $\nu_\mu$, and $\nu_\tau$,
respectively, and the off-diagonal terms encode phase coherence in the
$\mu$-$\tau$ sector.

Because each neutrino is produced at a different location and follows a
different trajectory through the complex inner region, the phases
$\varphi_k$ are effectively random. We assume that, in some basis in the
$\mu$-$\tau$ subspace, the sum of the interference terms averages to zero, so
that the ensemble can be treated as an incoherent mixture of flavor eigenstates.
Even if destructive interference happens in a rotated basis rather than in the
flavor basis, one can show that the off-diagonal element in the flavor basis is
typically small. The probability distribution of the relevant combination of
parameters is sharply peaked around zero, so that
$|\rho_{\mu\tau}|\ll1$ in most realizations. A detailed quantitative
argument is given in the Appendix of Ref.~\cite{Capanema:2025}.

Under this assumption we may summarize the ensemble at the boundary by the
flavor fractions
\begin{equation}
(a,b,c)_{\rm SN} \equiv
\bigl(f_{\nu_e}^{\rm SN},f_{\nu_\mu}^{\rm SN},f_{\nu_\tau}^{\rm SN}\bigr)
= \bigl(\langle \alpha^2\rangle,\langle \beta^2\rangle,\langle \gamma^2\rangle\bigr)
\end{equation}
where $\langle X\rangle \equiv \frac{1}{N}\sum_{k=1}^N X_k$ denotes an average
over the neutrino ensemble. We then follow their mapping to Earth through the
MSW region.

\begin{figure*}[t]
  \centering
  \includegraphics[width=\textwidth]{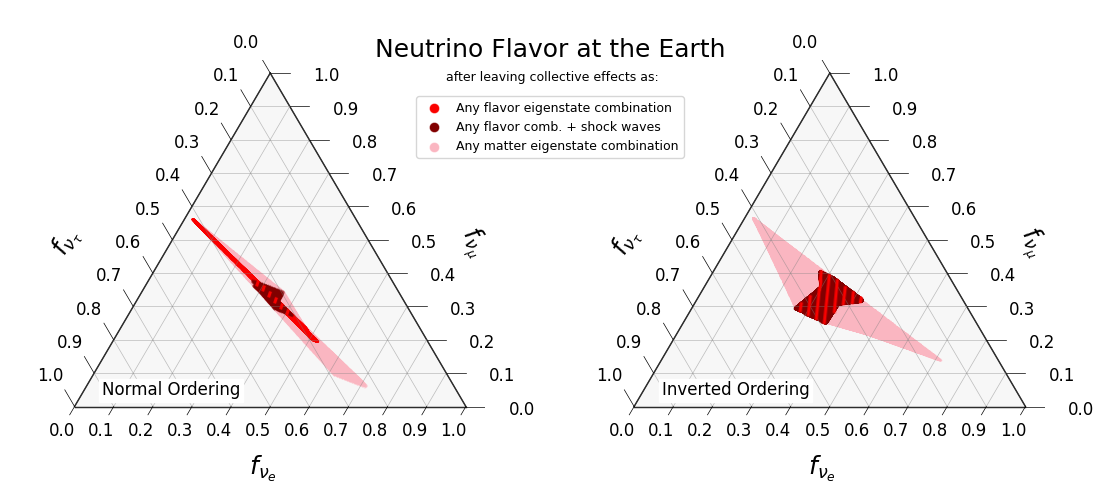}
  \caption{Allowed flavor compositions of supernova neutrinos at Earth for
  normal and inverted mass ordering. Red regions correspond to adiabatic
  propagation; maroon regions include possible nonadiabatic effects at the
  H resonance caused by shock waves; pink regions illustrate the larger space
  of possibilities if the ensemble emerges as an incoherent mixture of matter
  eigenstates. Figure adapted from Ref.~\cite{Capanema:2025}}
  \label{fig:triangle}
\end{figure*}

\section{Flavor fractions at Earth}

\subsection{Adiabatic propagation}

Assuming normal ordering and adiabatic propagation, an electron neutrino produced
above the H resonance exits the star as the third mass eigenstate, so that
\begin{equation}
(1,0,0)_{\rm SN} \rightarrow
\bigl(|U_{e3}|^2,|U_{\mu3}|^2,|U_{\tau3}|^2\bigr)_{\oplus}.
\end{equation}
A non–electron flavor at production emerges as an almost equal mixture of the
first and second mass eigenstates, which leads to
\begin{equation}
(0,1,0)_{\rm SN}~\text{or}~(0,0,1)_{\rm SN}
\rightarrow
\frac{1}{2}\bigl(|U_{e1}|^2+|U_{e2}|^2,\ldots\bigr)_{\oplus}.
\end{equation}
For a generic initial combination $(a,b,c)_{\rm SN}$ the electron-flavor
fraction at Earth is therefore
\begin{equation}
f_{\nu_e}^{\rm NO} \simeq a\,|U_{e3}|^2
+ \frac{b+c}{2}\bigl(|U_{e1}|^2+|U_{e2}|^2\bigr).
\end{equation}
Using unitarity, $a+b+c=1$, this simplifies to
\begin{equation}
f_{\nu_e}^{\rm NO} \simeq
\frac{1}{2}\bigl(1-|U_{e3}|^2\bigr)
+ \frac{a}{2}\bigl(3|U_{e3}|^2-1\bigr).
\end{equation}
Adopting $|U_{e3}|^2\simeq0.02$ and allowing $0\le a\le1$, we obtain the
general bound
\begin{equation}
f_{\nu_e}^{\rm NO} \simeq \frac{1-a}{2} \lesssim 0.5,
\end{equation}
with a small theoretical uncertainty due to the residual off-diagonal terms,
which in practice cannot be distinguished from the uncertainty associated with
the present errors on the oscillation parameters.

For inverted ordering the same reasoning applies with $U_{e3}\to U_{e2}$.
Since $|U_{e2}|^2\simeq 1/3$, the dependence on $a$ largely cancels and we
obtain
\begin{equation}
f_{\nu_e}^{\rm IO} \simeq \frac{1}{3},
\end{equation}
again with a modest uncertainty. Thus, under our decoherence assumption and
adiabatic evolution, the flavor content at Earth is tightly constrained for both
mass orderings.

\subsection{Effect of shock waves}

During the cooling phase, shock waves launched from the proto–neutron star can
propagate through the H-resonance region and temporarily spoil adiabaticity.
If an odd number of H crossings is nonadiabatic, transitions between the
relevant matter eigenstates effectively interchange the roles of $U_{e3}$ and
$U_{e2}$ in the expressions above. In NO this drives $f_{\nu_e}^{\rm NO}$ toward
$f_{\nu_e}\simeq1/3$, i.e.\ approximate flavor equipartition over the affected
energy range. In IO the H resonance is absent in the neutrino channel, so the
result remains $f_{\nu_e}^{\rm IO}\simeq1/3$.

The overall allowed regions in $(f_{\nu_e},f_{\nu_\mu},f_{\nu_\tau})$ at Earth
for the different scenarios are summarized in Fig.~\ref{fig:triangle}, where we
vary the relevant mixing angles over their current $3\sigma$ ranges.

\section{Discussion and outlook}

The main message of this work is that, despite our limited understanding of
collective oscillations in the inner core, the observable flavor composition of
SN neutrinos at Earth is not arbitrary. Under the plausible assumption that the
ensemble decoheres before entering the MSW region, standard matter effects
impose robust constraints: $f_{\nu_e}^{\rm NO}\lesssim0.5$ at all times and
energies, while IO generically leads to near flavor equipartition,
$f_{\nu_e}^{\rm IO}\simeq1/3$. Shock-induced nonadiabaticity at the H
resonance tends to erase the dependence on the initial flavor content and to
drive the system toward equipartition in both orderings. For antineutrinos the
constraints are somewhat weaker: in NO we find
$1/6\lesssim f_{\bar\nu_e}^{\rm NO}\lesssim0.7$, while in IO the adiabatic case
yields $f_{\bar\nu_e}^{\rm IO}\lesssim0.5$ and nonadiabatic H-resonance
crossings broaden the range to $1/6\lesssim f_{\bar\nu_e}^{\rm IO}\lesssim0.7$.
If the assumptions discussed above are not realized in Nature, a more
conservative bound can still be obtained: independently of the detailed flavor
evolution, both neutrinos and antineutrinos satisfy
$f_{\nu_e},f_{\bar\nu_e}\lesssim0.7$, corresponding to the pink regions in
Fig.~\ref{fig:triangle}. All these numerical ranges are derived in detail in
Ref.~\cite{Capanema:2025}.

A high-statistics observation of a future Galactic supernova would provide a
unique opportunity to confront these predictions with data from large neutrino
detectors sensitive to different interaction channels and energy ranges. Deviations from the bounds
derived here would signal either a failure of the decoherence assumption or the
presence of new physics affecting flavor conversion in the outer layers.

\begin{acknowledgments}
We thank the organizers of the 22nd Lomonosov Conference on Elementary Particle Physics for the invitation and an inspiring meeting.
\end{acknowledgments}


\end{document}